\documentclass[10pt]{article}
\usepackage[dvips]{graphicx,psfrag}
\usepackage{amssymb}
\usepackage{color}
\input epsf
\definecolor{Blue}{rgb}{0.3,0.3,0.9}
\definecolor{Red}{rgb}{1,0,0}
\definecolor{Green}{rgb}{0,1,0}
\newcommand{\be}{\begin{equation}}
\newcommand{\ee}{\end{equation}}
\newcommand{\bea}{\begin{eqnarray}}
\newcommand{\eea}{\end{eqnarray}}

\begin{document}
\begin{titlepage}

\begin{center}
{\Large\bf {Semiclassical scalar propagators in curved
backgrounds~:\\ formalism and ambiguities}}

\bigskip \bigskip \medskip
{\large J. Grain$^{1,2}$ \& A. Barrau$^1$
}\\[.5cm]
{\it $^1$Laboratoire de Physique Subatomique et de Cosmologie, Grenoble Universit\'es, CNRS, IN2P3\\ 
53, avenue de Martyrs, 38026 Grenoble cedex, France }\\[.5cm]
{\it $^2$AstroParticule \& Cosmologie, Universit\'e Paris 7, CNRS, IN2P3 \\
10, rue Alice Domon et L\'eonie Duquet, 75205 Paris cedex 13, France }\\[1cm]

{\bf Abstract}
\end{center}
The phenomenology of quantum systems in curved space-times is among the most 
fascinating fields of physics, allowing --often at the {\it gedankenexperiment}
level-- constraints on tentative theories of quantum gravity. Determining the 
dynamics of fields in curved backgrounds remains however a complicated 
task because of the highly intricate partial differential equations involved,
especially when the space metric exhibits no symmetry. In 
this article, we provide --in a pedagogical way-- a general formalism to 
determine this dynamics at the 
semiclassical order. To this purpose, a generic expression for the 
semiclassical propagator is computed and the equation of motion for the 
probability four-current is derived. Those results underline a direct analogy 
between the computation of the propagator in general relativistic quantum 
mechanics and the computation of the propagator for stationary systems in 
non-relativistic quantum mechanics. A possible application of this formalism 
to curvature-induced quantum interferences is also discussed.
\\[1cm]

PACS numbers: 04.62.+v, 11.15.Kc
\end{titlepage}

\section{Introduction}

	The dynamics of a scalar field propagating in a curved background is governed by partial 
	differential equations which, in
most cases, have no analytical solution. Investigating the behavior of those fields in the 
semiclassical approximation is a promising alternative to numerical studies, allowing accurate predictions for many phenomena including the Hawking radiation process
\cite{hawking,padmanabhan,parentani,parikh} and the primordial power spectrum \cite{martin,casadio}. We
have determined the 
formalism providing the
general expression of the WKB wave function in any static and spherically symmetric backgrounds in
\cite{wkb}. In this article, the aforementioned study is extended to the general case of scalar fields evolving in any curved
background. To the best of our knowledge, such a general formalism has not yet been published, although some WKB solutions were derived in
the small proper time limit \cite{dewitt} or for non-relativistic particles in curved spaces
\cite{dewitt1}.

	After a brief reminder of the classical dynamics for scalar particles in Section \ref{sec0}, the propagator is built, following
\cite{feynman,hartle}, using an action functional quadratic in four-velocity and completely defined up to a
multiplicative constant $\alpha$. However, the definition used hereafter is also inspired by non-relativistic semiclassical quantum
mechanics, leading to some discrepancies with the previous definition. Section
\ref{sec1} is devoted to the computation of the semiclassical propagator which is performed in
two steps~: the derivation of the propagator for paths at fixed proper times and its translation into a propagator at fixed
masses. This computation is performed in a four-dimensional framework and then extended to the $D-$dimensional case. The equation of motion is then derived in Section \ref{sec2}, allowing the determination of the prefactor in the action by requiring
the propagator to be a solution of the inhomogeneous Klein-Gordon (KG) equation. We show in this section that the product of
the square modulus of the semiclassical amplitude by the classical four-momentum can be interpreted as a four-current at
the semiclassical order, just as in non-relativistic quantum mechanics. The domain of validity of the semiclassical
approximation is also drawn out in Section \ref{sec2}. We discuss our results in Section \ref{sec3} in the light of the non-relativistic semiclassical propagator and compare them with previously obtained solutions \cite{dewitt,dewitt1}. The choice of the action functional is also discussed in this part. 
In Section \ref{interference}, we suggest a possible experiment based on 
gravitationnally-induced quantum interferences which could help to clearly 
define propagators in curved backgrounds. To the best of our knowledge, it is 
the first time that quantum interferences due to space-time curvature are 
considered in the literature to this purpose. Finally, our conclusions are summarized in the last section of this
article
\footnote{In this article, we choose the $(+---)$ signature for the metric tensor. It should be noticed 
that all the results mentioned hereafter are rewritten from the original articles with this convention 
although most of them --especially in \cite{hartle}-- were originally derived with the $(-+++)$ convention.}.

\section{Lagrangian mechanics in curved spaces and definition of the propagator}
\label{sec0}
	For a particle moving in a $4$-dimensional curved background, the action can be written as a 
	functional of the general coordinate system $\left[x^\mu(\tau)\right]$. Each coordinate is a function of an affine parameter $\tau$ which can be chosen to be the proper time for massive
	 particles. The action functional is given by~:
\begin{equation}
	S=m\displaystyle\int^{\tau}_0\sqrt{g_{\mu\nu}\dot{x}^\mu\dot{x}^\nu}{d\tau},
\end{equation}
where the $\dot{x}^\mu$ notation stands for the derivative according to the proper time and $m$ is the mass of the particle. Nevertheless,
as heuristically mentioned in \cite{hartle}, a Lagrangian quadratic in four-velocity should be preferred for the study of path integrals. This choice is of course partially motivated by the analogy with classical mechanics. We will therefore use the following action functional~:
\begin{equation}
	S=\alpha\displaystyle\int^{\tau}_0g_{\mu\nu}\dot{x}^\mu\dot{x}^\nu{d\tau},
\end{equation}
where $\alpha$ is a constant to be determined later. Although both definitions 
are classically valid as they lead to the same geodesic 
equation, the difference between the two formulations at the quantum level will be discussed in Section \ref{sec3}. 
Along a trajectory described by the classical action $\tilde{S}$, the four-momentum is defined
 by~:
\begin{equation}
	p_\mu=\frac{\partial\tilde{S}}{\partial{x}^\mu}=2\alpha\dot{x}_\mu,
\end{equation}
and the action satisfies the following Hamilton-Jacobi (HJ) formula~:
\begin{equation}
	g^{\mu\nu}p_\mu{p}_\nu=4\alpha^2.
	\label{H-J-momentum}
\end{equation}
Differentiating the action according to the proper time gives $-\frac{\partial\tilde{S}}{\partial\tau}=\alpha$, leading to the HJ equation expressed with partial derivatives of the classical action~:
\begin{equation}
	\left(\frac{\partial\tilde{S}}{\partial\tau}\right)^2=\frac{1}{4}g^{\mu\nu}\frac{\partial\tilde{S}}{\partial{x}^\mu}\frac{\partial\tilde{S}}{\partial{x}^\nu}.
	\label{H-J-action}
\end{equation}
>From the above remarks, $\alpha$ can clearly be interpreted as the conjugate variable of the proper time while the four-momentum $p_\mu$ is the conjugate variable of the coordinates $x^\mu$.

	In quantum mechanics, the propagator can be defined as the sum over all paths joining one event to the other, the exponential of the action along each path giving the probability amplitude~:
\begin{equation}
	K\left[x^\mu(\tau),x^\mu(0)\right]=\displaystyle\int^{x^\mu(\tau)}_{x^\mu(0)}\mathcal{D}\left[x^\mu(\tau')\right]e^{\frac{i}{\hbar}S\left[x^\mu(\tau'),x^\mu(0)\right]}.
\end{equation}
In the following, the $0$ subscript means that the quantity is evaluated at the initial proper time. In a relativistic framework, the
proper time is not observed and the propagator given above has to be averaged
\cite{hartle,feynman}. The following form should consequently be considered~:
\begin{equation}
	G\left[x^\mu,x^\mu(0),\alpha\right]=i\displaystyle\int^\infty_0d\tau{e}^{\frac{i}{\hbar}\alpha\tau}K\left[x^\mu(\tau),x^\mu(0)\right].
	\label{complete-prop}
\end{equation}
This transformation is reminiscent to the case of stationary systems in non-relativistic quantum mechanics~: the above Fourier
transform in the proper time domain allows us to consider paths with a fixed value of $\alpha$ instead of paths with fixed proper times,
just like the Fourier transform in the frequency domain in non-relativistic quantum mechanics translates paths with fixed times into
paths with fixed energies. The $\alpha$ constant should therefore be an observable quantity characterizing the considered particle
(typically its mass or something directly related with its mass).

	It is important to mention that this second step in the definition of the propagator is slightly different from the one used in \cite{hartle}~: although the motivation is the same in both definitions (changing the {\it{unobservable}} proper time to the {\it{observable}} mass), the chosen weight in our study differs from \cite{hartle} as our main guideline is to build the semiclassical solution and therefore take advantage of the stationary phases
approximation. This question will be addressed in the Section \ref{sec3} of this note. 

\section{Propagator at the semiclassical order}
\label{sec1}
	\subsection{Proper-time semiclassical propagator}

	The semiclassical approximation consists in taking into account only the first quantum 
	correction around the classical trajectory. To this purpose, we perform a Taylor expansion of the action around the classical path $\left[\bar{x}^\mu(\tau')\right]$~:
\begin{equation}
	S\left[(\bar{x}^\mu+\delta{x}^\mu)(\tau'),x^\mu(0)\right]\simeq{S}\left[\bar{x}^\mu(\tau'),x^\mu(0)\right]+\frac{1}{2}\left.\frac{\delta^2S}{\delta{x}^2}\right|_{\bar{x}^\mu(\tau')}\delta{x}^2(\tau').
\end{equation}
The first-order term in this expansion is directly set equal to zero as it
corresponds to the least action principle and the zeroth order term is the
classical action $\tilde{S}$. In this approximation scheme, the $K$ propagator
now reads~:
\begin{equation}
	\tilde{K}\left[x^\mu(\tau),x^\mu(0)\right]=F\left[x^\mu(\tau),x^\mu(0)\right]e^{\frac{i}{\hbar}\tilde{S}\left[x^\mu(\tau'),x^\mu(0)\right]},
\end{equation}
where all the first order quantum corrections are contained into the $F$ functional~:
\begin{equation}
	F\left[x^\mu(\tau),x^\mu(0)\right]=\displaystyle\int^{0,\tau}_{0,0}\mathcal{D}\left[\delta{x}^\mu(\tau')\right]\exp{\left(\frac{i}{2\hbar}\left.\frac{\delta^2S}{\delta{x}^2}\right|_{\bar{x}^\mu(\tau')}\delta{x}^2(\tau')\right)}
	\label{f-gen}
\end{equation}
The above functional can be determined thanks to two properties characterizing the propagator. 

\begin{flushleft}
{\underline{Square modulus of the $F$ functional~:}}
\end{flushleft}
	Following \cite{morette}, we determine the square modulus of $F$ thanks to 
	{\it unitary conditions}~: the probability to propagate from a given initial point to 
	{\it{any}} point is equal to one. Determining $\left|F\right|^2$ is equivalent to the 
	computation of the normalization constant of the propagator during an infinitesimally small 
	interval (as in this case the dominant paths are the classical ones leading to a normalization
	 constant equal to the inverse of $F$). For an infinitesimally small proper time interval 
	 $\varepsilon$, the unitarity conditions can then be viewed as a two steps propagation starting 
	 from the initial point $x^\mu(0)$ to the final one $x^\mu(\varepsilon)$ followed by the return,
	  back in proper time, from $x^\mu(\varepsilon)$ to $y^\mu(0)$, the intermediate point $x^\mu(\varepsilon)$ covering all the space-time~:
\begin{equation}
	\frac{\delta^4\left(x^\mu(0)-y^\mu(0)\right)}{\sqrt{-g(0)}}=\displaystyle\int\mathcal{D}\left[x^\mu(\varepsilon)\right]K^\dag\left[x^\mu(\varepsilon),y^\mu(0)\right]{K}\left[x^\mu(\varepsilon),x^\mu(0)\right].
\end{equation}
The determinant of the metric tensor appears in the left-hand-side of the above
equation as the Dirac distribution is a tensor
density with weight -1 ($\delta\left[f(x)\right]\equiv\left|\frac{df}{dx}\right|^{-1}\delta(x)$). For the above relations to fully describe
the unitary conditions, $y^\mu(0)$ should lie within an infinitesimally small interval around $x^\mu(0)$. It
should be notice that this integral is not over a path but over the whole space-time covered by the
$x^\mu(\varepsilon)$ coordinates leading to
$\mathcal{D}\left[x^\mu(\varepsilon)\right]=\sqrt{-g(\varepsilon)}d^4x(\varepsilon)$. The classical action can be expanded
in the vicinity of the intermediate event $x^\mu(\varepsilon)$ \footnote{The Taylor expansion should in general be performed using Lie
derivatives. However, the action being a scalar, the Lie derivatives reduce to partial derivatives.}~:
\begin{equation}
	\tilde{S}\left[x^\mu(\varepsilon),x^\mu(0)\right]\simeq\tilde{S}\left[x^\mu(\varepsilon),x^\mu(\varepsilon)\right]+g_{\mu\nu}p^\nu(0)\left(x^\mu(0)-x^\mu(\varepsilon)\right),
\end{equation}
and
\begin{equation}
	\tilde{S}\left[x^\mu(\varepsilon),y^\mu(0)\right]\simeq\tilde{S}\left[x^\mu(\varepsilon),x^\mu(\varepsilon)\right]+g_{\mu\nu}p^\nu(0)\left(y^\mu(0)-x^\mu(\varepsilon)\right).
\end{equation}
The four-vector ${p_\mu}(0)$ stands for the derivative of the action with respect to the coordinate system at the initial time. In the language of classical mechanics, this corresponds to the classical four-momentum evaluated at the initial proper time which is the same for the two paths $x^\mu(0)\to{x}^\mu(\varepsilon)$ and $y^\mu(0)\to{x}^\mu(\varepsilon)$ 
at this level of approximation. We finally obtain the following conditions~:
\begin{equation}
	\frac{\delta^4\left(x^\mu(0)-y^\mu(0)\right)}{\sqrt{-g(0)}}=\displaystyle\int{d}^4x(\varepsilon)\sqrt{-g(\varepsilon)}\left|F\right|^2\exp{\left[\frac{i}{\hbar}g_{\mu\nu}p^\nu(0)\left(x^\mu(0)-y^\mu(0)\right)\right]}.
	\label{unitarity}
\end{equation}
As for an infinitesimally small propagation, the propagator is dominated by classical paths ({\it{i.e.}} stationary paths), the intermediate 
point is totally defined by the initial classical four-momentum and the integral over $x^\mu(\varepsilon)$ 
can be changed into an integral over $p^\nu(0)$~:
\begin{eqnarray}
	x^\mu(\varepsilon)&\rightarrow&g_{\mu\nu}p^\nu(0), \\
	d^4x(\varepsilon)&\rightarrow&\frac{\partial{x^0}}{\partial{{p_\mu}(0)}}d{p_\mu}(0)\frac{\partial{x^1}}{\partial{{p_\nu}(0)}}d{p_\nu}(0)\frac{\partial{x^2}}{\partial{{p_\alpha}(0)}}d{p_\alpha}(0)\frac{\partial{x^4}}{\partial{{p_\beta}(0)}}d{p_\beta}(0).
\end{eqnarray}
The differential $4-$volume element can be written using the determinant~:
\begin{equation}
	d^4x(\varepsilon)=C\left|\det_{\mu\nu}\left[\frac{\partial{p_\nu}(0)}{\partial{x}^\mu(\varepsilon)}\right]\right|^{-1}{dp_0}(0){dp_1}(0){dp_2}(0){dp_3}(0).
	\label{det-cste}
\end{equation}
The quantities $d^4x^\mu(\varepsilon)$, $d^4{p_\nu}(0)$ and
$\left|\det_{\mu\nu}\left[\frac{\partial{p_\nu}(0)}{\partial{x}^\mu(\varepsilon)}\right]\right|^{-1}\equiv\left|\tilde{S}^{\mu\nu}_{,\mu\nu}\right|$
are tensor densities with respective weights 1 at $\varepsilon$, -1 at $0$ and $2$,
distributed between the initial and final events. Through a global change of
coordinates $x^\mu\rightarrow{y}^\alpha$, they are modified in the following
way~:
\begin{eqnarray}
	d^4x^\mu&\rightarrow&d^4y^\alpha=\left|\frac{d\mathbf{Y}}{d\mathbf{X}}\right|d^4x^\mu, \\
	d^4{p_\mu}(0)&\rightarrow&d^4{q_\alpha}(0)=\left|\frac{d\mathbf{Y}(0)}{d\mathbf{X}(0)}\right|^{-1}d^4{p_\mu}(0), \\
	\left|\tilde{S}^{\mu\nu}_{,\mu\nu}\right|&\rightarrow&\left|\tilde{S}^{\alpha\beta}_{,\alpha\beta}\right|=\left|\frac{d\mathbf{Y}}{d\mathbf{X}}\right|\left|\frac{d\mathbf{Y}(0)}{d\mathbf{X}(0)}\right|\left|\tilde{S}^{\mu\nu}_{,\mu\nu}\right|.
\end{eqnarray}
The notation $\frac{d\mathbf{Y}}{d\mathbf{X}}$ stands for the determinant of the Jacobian. To insure 
that Eq. (\ref{det-cste}) is satisfied, the constant $C$ should be taken equal to one. For the two sides of Eq. (\ref{unitarity}) to be equal, the square modulus of the $F$ functional has to be set to~:
\begin{equation}
	\left|F\right|^2\left[x^\mu(\tau),x^\mu(0)\right]=\frac{1}{(2\pi\hbar)^4\sqrt{g(\tau)g(0)}}\times\left|\det_{\mu\nu}\left[\frac{\partial^2\tilde{S}}{\partial{x}^\mu(\tau)\partial{x}^\nu(0)}\right]\right|.
\end{equation}
The above results are the natural extension to curved spaces of the square
modulus of the semiclassical amplitude derived in \cite{morette} in the case of
special relativistic quantum mechanics~: the additional prefactor
$(g(\tau)g(0))^{-1/2}$ insures that the $\left|F\right|^2$ functional will be a scalar.

\begin{flushleft}
{\underline{Phase of the $F$ functionnal~:}}
\end{flushleft}
	
	The computation of the phase depends on the nature of the point of the
 classical trajectory~: {\it{regular}} point, in the sense that the momentum is 
not zero, and {\it{turning}} or {\it{focal}} points, where $p^\mu$ vanishes. 
Just as in non-relativistic quantum mechanics, we expect some particular 
effects, leading to an additional global phase, to occur at those focal and 
turning points. Computing accurately this phase at turning points requires to 
use the Morse theory and is beyond the scope of this work. We will however give 
some heuristic argument to determine the value of the phase.

	The phase at regular points is determined using the semigroup properties of the propagator
	\cite{morette}. Quantum mechanics being a Markovian theory, the propagator, interpreted as probability amplitude, satisfies the condition~:
\begin{equation}
	K\left[x^\mu(\tau),x^\mu(0)\right]=\displaystyle\int\mathcal{D}\left[x^\mu(\tau')\right]K\left[x^\mu(\tau),x^\mu(\tau')\right]{K}\left[x^\mu(\tau'),x^\mu(0)\right],
\end{equation}
and therefore belongs to a semigroup. This statement should be conserved at the semiclassical order, 
which allows us to derive the phase of the $F$ functional. For this purpose, we consider a infinitesimal 
path from $\tau=0$ to $\tau=2\varepsilon$, divided into two successive paths, from $x^\mu(0)$ at 
$\tau=0$ to $x^\mu(\varepsilon)$ at $\tau=\varepsilon$ and then from $x^\mu(\varepsilon)$ at 
$\tau=\varepsilon$ to $x^\mu(f)$ at $\tau=2\varepsilon$. The action along these classical paths is
 expressed using a Riemann normal coordinate system defined by 
 $z^\mu_{\tau_1\to\tau_2}=x^\mu(\tau_2)-x^\mu(\tau_1)$, in which the geodesics, during an infinitesimal 
 proper time, become straight lines and the metric tensor tends to the Minkowski one at the first order. This approximation method is valid as we are considering infinitesimally small paths, providing classical actions of the form~:
\begin{eqnarray}
	\tilde{S}_{0\to\varepsilon}\equiv\tilde{S}[x^\mu(\varepsilon),x^\mu(0)]&=&\frac{\alpha}{\varepsilon}\eta_{\mu\nu}z^\mu_{0\to\varepsilon}z^\nu_{0\to\varepsilon}, \\
	\tilde{S}_{\varepsilon\to2\varepsilon}\equiv\tilde{S}[x^\mu(f),x^\mu(\varepsilon)]&=&\frac{\alpha}{\varepsilon}\eta_{\mu\nu}z^\mu_{\varepsilon\to2\varepsilon}z^\nu_{\varepsilon\to2\varepsilon},
\end{eqnarray}
and
\begin{equation}
	\tilde{S}_{0\to2\varepsilon}\equiv\tilde{S}[x^\mu(f),x^\mu(0)]=\frac{\alpha}{2\varepsilon}\eta_{\mu\nu}z^\mu_{0\to2\varepsilon}z^\nu_{0\to2\varepsilon}.
\end{equation}
>From these actions and the definition of the square modulus of the functional, 
the semigroup 
conditions can be written as~:
\begin{equation}
	\gamma{e}^{-i\varphi}{e}^{\frac{i}{\hbar}\tilde{S}_{0\to2\varepsilon}}=\underbrace{\displaystyle\int{d}^4x(\varepsilon)e^{\frac{i}{\hbar}\left(\tilde{S}_{0\to\varepsilon}+\tilde{S}_{\varepsilon\to2\varepsilon}\right)}}_{I_{0\to2\varepsilon}},
	\label{phase-int}
\end{equation}
with
\begin{equation}
	\gamma=\left(\frac{\pi\hbar\varepsilon}{2\alpha}\right)^2
\end{equation}
and $\varphi$ corresponds to the phase of the $F$ functional. It is important to notice that in Eq. (\ref{phase-int}) all the $\sqrt{-g(\tau)}$ cancel out because of the $\left(g(\tau)g(0)\right)^{-1/4}$ prefactor in the $F$ functional. As a consequence, the above integral reduces to the convolution of two Gaussian functions with variance $\sigma^2=i\hbar\varepsilon/2\alpha$ per degree of freedom~:
\begin{eqnarray}
	I_{0\to2\varepsilon}&=&\left(\sigma\sqrt{\pi}\right)^4\exp{\left(\frac{i\alpha}{2\hbar\varepsilon}\eta_{\mu\nu}z^\mu_{0\to2\varepsilon}z^\nu_{0\to2\varepsilon}\right)}, \nonumber \\
	&=&\gamma{e}^{i\pi}e^{\frac{i}{\hbar}\tilde{S}_{0\to2\varepsilon}}.
\end{eqnarray}
The phase of the $F$ functional is consequently equal to $\varphi=-\pi$, just
 like in special relativistic quantum mechanics where it
was proved \cite{morette} that the phase is $-\pi/4$ per degree of freedom.

	The additional phase, appearing at turning points, has been computed in a non-relativistic
framework \cite{schulman} as well as for the propagation in Schwarzschild-like spaces \cite{wkb} and was
shown to be equal to $-\pi/2$ at each of these
particular points. Unfortunately, the semiclassical approximation breaks down at
focal and turning points because the momentum vanishes. As a consequence, the WKB propagator diverges 
and the computation of the
additional phases requires a careful treatment. Nevertheless, it can be noticed that at turning points, 
the sign of the momentum changes and,
from Eq. (\ref{f-func}), the $F$ functional should acquire a $\pm\pi/2$ 
additional phase. Although the sign cannot be determined in this framework, it 
should be kept in mind that this will have no consequence for the rest of this work.

	This computation ends the derivation of the semiclassical {\it{proper-time}} propagator
	$\tilde{K}$ which takes the following form~:
\begin{equation}
	\tilde{K}\left[x^\mu(\tau),x^\mu(0)\right]=F\left[x^\mu(\tau),x^\mu(0)\right]\exp{\left(\frac{i}{\hbar}\tilde{S}\left[x^\mu(\tau),x^\mu(0)\right]\pm{i}q\frac{\pi}{2}\right)},
\end{equation}
where
\begin{equation}
	\tilde{S}\left[x^\mu(\tau),x^\mu(0)\right]=\alpha\displaystyle\int^{\tau}_0{g}_{\mu\nu}\dot{x}^\mu\dot{x}^\nu{d\bar{\tau}},
\end{equation}
and $q$ is the number of focal and turning points along the classical trajectory. The $F$ functional is given by a Van
Wleck-Morette determinant~:
\begin{equation}
	F\left[x^\mu(\tau),x^\mu(0)\right]=\left\{g(\tau)g(0)\right\}^{-1/4}\times\left\{\det_{\mu\nu}\left[(2i\pi\hbar)^{-1}\frac{\partial^2\tilde{S}}{\partial{x}^\mu\partial{x}^\nu(0)}\right]\right\}^{1/2}. \nonumber
	\label{f-func}
\end{equation}
In the definition of the classical action, the integration is performed along a classical path, denoted by {\it{barred}} quantities, and the upper bound as to be thought of as the classical proper time at which the particle reaches its final space-time position.

	\subsection{Complete semiclassical propagator}

	In order to simplify the formula, the additional phases at focal and turning points will be omitted hereafter as they are not involved in the
computation of the complete propagator.

	The {\it{proper-time}} semiclassical propagator must now be introduced in the definition of the complete propagator $G$ and the integral over the proper time must be performed. However, at the semiclassical order, this integral is greatly simplified~: the
semiclassical propagator is dominated by stationary paths which correspond to only one proper time $\bar{\tau}$ fixed by the geodesic
equation \footnote{In principle, the geodesic equation can lead to different classical proper times $\bar{\tau}$ (periodic motions are
good examples). The complete solution is given by summing over all the classical trajectories, taking into account the number of focal and
turning points for each classical path.}. Therefore, the complete propagator is given by a Taylor expansion around the classical proper time\footnote{This is precisely the {\it stationary phase approximation}.}~:
\begin{equation}
\tilde{G}\left[x^\mu,x^\mu(0),\alpha\right]=i\displaystyle\int^\infty_0F\left[x^\mu(\tau),x^\mu(0)\right]\exp{\left(\frac{i}{\hbar}\varphi\left[x^\mu(\tau),x^\mu(0)\right]\right)}d\tau, \nonumber
\end{equation}
with the phase
\begin{equation}
	\varphi\left[x^\mu(\tau),x^\mu(0)\right]\approx\alpha\tau+\tilde{S}\left[x^\mu(\bar{\tau}),x^\mu(0)\right]+\left.\frac{\partial\tilde{S}}{\partial\tau}\right|_{\bar{\tau}}\left(\tau-\bar{\tau}\right)+\frac{1}{2}\left.\frac{\partial^2\tilde{S}}{\partial\tau^2}\right|_{\bar{\tau}}\left(\tau-\bar{\tau}\right)^2.
\end{equation}
Dealing with classical paths, the action at $\bar{\tau}$ and its first partial derivative are given 
by~:
\begin{eqnarray}
	\frac{\partial\tilde{S}}{\partial\tau}&=&-\alpha, \\
	\tilde{S}\left[x^\mu(\bar{\tau}),x^\mu(0)\right]&=&\displaystyle\int^{x^\mu(\bar{\tau})}_{x^\mu(0)}\left(\frac{\partial\tilde{S}}{\partial\tau}d\tau+\frac{\partial\tilde{S}}{\partial{x}^\mu}dx^\mu\right).
\end{eqnarray}	
Replacing the partial derivatives by their expressions given in the introduction, the phase of the propagator is finally decomposed into a
zeroth order term, $\bar{\varphi}$, a vanishing first order term and a non-vanishing second order one, 
$\varphi^{(2)}$~:
\begin{eqnarray}
	\bar{\varphi}\left[x^\mu(\tau),x^\mu(0)\right]&=&\displaystyle\int^{x^\mu}_{x^\mu(0)}g_{\mu\nu}p^\nu{d}\bar{x}^\mu, \\
	\varphi^{(2)}\left[x^\mu(\tau),x^\mu(0)\right]&=&\frac{1}{2}\left.\frac{\partial^2\tilde{S}}{\partial\tau^2}\right|_{\bar{\tau}}\left(\tau-\bar{\tau}\right)^2,
\end{eqnarray}
where the $d\bar{x}^\mu$ means that the integration is performed along the classical geodesic joining
$x^\mu(0)$ to $x^\mu$. In the
above equations, $p_\mu{d}\bar{x}^\mu$ was re-written as $g_{\mu\nu}p^\nu{d}\bar{x}^\mu$ in order to underline the dependence upon the
background curvature.

	The complete propagator now appears as the product of an exponential, depending only on the 
	classical trajectory, with an integral over the  proper time~:
\begin{equation}
\tilde{G}\left[x^\mu,x^\mu(0),\alpha\right]=i\exp{\left(\frac{i}{\hbar}\displaystyle\int^{x^\mu}_{x^\mu(0)}g_{\mu\nu}p^\nu{d}\bar{x}^\mu\right)}\times\displaystyle\int^\infty_0{F}\left[x^\mu(\bar{\tau}),x^\mu(0)\right]e^{\frac{i}{2\hbar}\left.\frac{\partial^2\tilde{S}}{\partial\tau^2}\right|_{\bar{\tau}}\left(\tau-\bar{\tau}\right)^2}d\tau. \nonumber
\end{equation}
The $F$ functional is not involved in the proper time integral as it only depends upon the authorized classical time $\bar{\tau}$
 as the first order contribution vanishes\footnote{Unlike in special relativity,
 $\left.\frac{\partial{F}}{\partial\tau}\right|_{\bar{\tau}}$ is not equal to zero because of the variation of the metric tensor. However,
 performing a Taylor expansion of the $F$ functional around $\bar{\tau}$ leads us to evaluate the additional integral
$I^{(1)}=\left.\frac{\partial{F}}{\partial\tau}\right|_{\bar{\tau}}\displaystyle\int(\tau-\bar{\tau})\times\exp{(\varphi^{(2)})}d\tau$.
This integral is equal to zero as it corresponds to the expectation value of a gaussian and centered random variable.}. Consequently, the integrand is a Gaussian function with a variance given by the inverse of the second derivative of $\tilde{S}$. The complete propagator is finally written as~:
\begin{equation}
	\tilde{G}\left[x^\mu,x^\mu(0),\alpha\right]=i{F}\left[x^\mu(\bar{\tau}),x^\mu(0)\right]\sqrt{\frac{2i\pi\hbar}{\left.\frac{\partial^2\tilde{S}}{\partial\tau^2}\right|_{\bar{\tau}}}}\times\exp{\left(\frac{i}{\hbar}\displaystyle\int^{x^\mu}_{x^\mu(0)}g_{\mu\nu}p^\nu{d}\bar{x}^\mu-iq\frac{\pi}{2}\right)},
\end{equation}
where we have re-introduce the phases at focal and turning points for completeness. From the HJ equation (\ref{H-J-action}), the second derivative of the classical action can be written as a function of the 
four-velocity and four momentum~:
\begin{equation}
	\left.\frac{\partial^2\tilde{S}}{\partial\tau^2}\right|_{\bar{\tau}}=\frac{1}{2}g^{\mu\nu}\left(\frac{\partial\bar{p}_\mu}{\partial\tau}\right)\left(\frac{\partial\bar{x}_\nu}{\partial\tau}\right),
\end{equation}
where the {\it{barred}} quantities mean that they are evaluated at the classical proper time $\bar{\tau}$. In addition, the determinant involved in the $F$ functional reads as~:
\begin{equation}
	\det_{\mu\nu}{\left[\frac{\partial^2\tilde{S}}{\partial{x}^\mu\partial{x}^\nu(0)}\right]}=\det_{\mu\nu}{\left[\frac{\partial\bar{p}_\mu}{\partial\tau}\times\left(\frac{\partial{x}^\nu(0)}{\partial\tau}\right)^{-1}\right]}.
\end{equation}
The square modulus of the $\tilde{G}$ propagator is proportional to the ratio of the two quantities previously computed~:
\begin{equation}
	\left|\tilde{G}\right|^2\propto\left|\frac{\epsilon^{\beta\gamma\lambda\rho}\left(g_{\sigma\beta}\dot{p}^\sigma\right)\left(g_{\sigma\gamma}\dot{p}^\sigma\right)\left(g_{\sigma\lambda}\dot{p}^\sigma\right)\left(g_{\sigma\rho}\dot{p}^\sigma\right)}{\frac{1}{2}g_{\mu\nu}\dot{p}^\mu\dot{x}^\nu\displaystyle\prod^3_{\kappa=0}\dot{x}^\kappa(0)}\right|,
\end{equation}
where $\epsilon$ is the totally antisymmetric tensor. Calling $\mathcal{H}$ the right-hand side of the above equation, the
complete propagator finally reads as~:
\begin{equation}
	\tilde{G}\left[x^\mu,x^\mu(0),\alpha\right]=\frac{i\sqrt{\mathcal{H}\left[x^\mu(\bar{\tau}),x^\mu(0)\right]}}{\sqrt{\left(2i\pi\hbar\right)^3\sqrt{g(\bar{\tau})g(0)}}}\exp{\left(\frac{i}{\hbar}\displaystyle\int^{x^\mu}_{x^\mu(0)}g_{\mu\nu}p^\nu{d}\bar{x}^\mu-iq\frac{\pi}{2}\right)}.
\end{equation}
In this equation, the knowledge of the boundary conditions $\left(x^\mu,x^\mu(0)\right)$ is in principle sufficient to determine the propagator~: once the geodesic equation is solved, the proper time $\bar{\tau}$ as well as the classical action can be computed. However, it should be kept in mind that solving the classical problem might be very complicated.

	\subsection{$D-$dimensional generalization}
	The aforementioned semiclassical propagator has been computed in a four-dimensional framework. However, it can be easily
extended to the propagation in a $D-$dimensional background. Assuming that the
$D-4$ additional dimensions are of space type and are large, all the previous calculations are easily performed to give~:
\begin{equation}
	\tilde{G}\left[x^\mu,x^\mu(0),\alpha\right]=\frac{i\sqrt{\mathcal{H}\left[x^\mu(\bar{\tau}),x^\mu(0)\right]}}{\sqrt{\left(2i\pi\hbar\right)^{D-1}\sqrt{g(\bar{\tau})g(0)}}}\exp{\left(\frac{i}{\hbar}\displaystyle\int^{x^\mu}_{x^\mu(0)}g_{\mu\nu}p^\nu{d}\bar{x}^\mu-iq\frac{\pi}{2}\right)}, \nonumber
\end{equation}
where all the indices run from $0$ to $D-1$. In particular, the $\mathcal{H}$ function reads~:
\begin{equation}
	\mathcal{H}=\left|\frac{\epsilon^{\beta_0\cdots\beta_{D-1}}\left(g_{\sigma\beta_0}\dot{p}^{\sigma}\right)\cdots\left(g_{\sigma\beta_{D-1}}\dot{p}^{\sigma}\right)}{\frac{1}{2}g_{\mu\nu}\dot{p}^\mu\dot{x}^\nu\displaystyle\prod^{D-1}_{\kappa=0}\dot{x}^\kappa(0)}\right|.
\end{equation}
The number of dimensions therefore influences the propagator through both its phase and its amplitude.

\section{Semiclassical equation of motion}
\label{sec2}
	\subsection{Equation of motion and determination of $\alpha$}
	The equation of motion for the {\it{proper-time}} propagator has been investigated in \cite{dewitt,cheng,hartle} in a fully quantum mechanical framework. It was shown that the $K$ propagator should be a solution of a Schr\"odinger-like equation \cite{hartle}~:
\begin{equation}
	i\hbar\frac{\partial{K}}{\partial\tau}=-\hbar^2\left[\Box^2-\frac{1}{3}R\right]K,
	\label{schro-K1}
\end{equation}
where $\Box^2$ is the KG operator defined by 
$\Box^2=\frac{1}{\sqrt{-g}}\partial_\mu\sqrt{-g}g^{\mu\nu}\partial_\nu$ and $R$ stands for the Ricci 
scalar. As shown by \cite{dewitt1}, the quantum analog of the HJ equation should
be written replacing the standard Hamiltonian $H$ by the Hamiltonian operator
$\hat\mathcal{H}=\hat{H}+\frac{\hbar^2}{12}R$, $\hat{H}$ being the operator
associated with $H$. As a consequence, for an action functional of the form 
$S=\frac{1}{4}\displaystyle\int{d\tau}g_{\mu\nu}\dot{x}^\mu\dot{x}^\nu$, the 
Hamiltonian operator reads
$\hat\mathcal{H}=-\hbar^2\left[\Box^2-\frac{1}{3}R\right]$. This result
implicitly assumes that the Feynman propagator is decomposed as \cite{dewitt1}~:
\begin{equation}
	K\left[x^\mu(\tau),x^\mu(0)\right]=\lim_{N\to\infty}\displaystyle\prod^N_{i=1}\int\sqrt{-g(\varepsilon_i)}d^4x_iF\left[x^\mu_{i+1},x^\mu_i\right]\exp{\left(\frac{i}{\hbar}S\left[x^\mu_{i+1},x^\mu_i\right]\right)}.
\end{equation}
However, as mentioned in \cite{hartle}, the amount of scalar curvature in Eq. (\ref{schro-K1}) can be set equal to zero if $\exp{\left(\frac{i}{\hbar}S\left[x^\mu_{i+1},x^\mu_i\right]\right)}$ is replaced by $\sqrt{\frac{g(\varepsilon_{i+1})}{g(\varepsilon_{i})}}\exp{\left(\frac{i}{\hbar}S\left[x^\mu_{i+1},x^\mu_i\right]\right)}$, leading to the following dynamical equation~:
\begin{equation}
	i\hbar\frac{\partial{K}}{\partial\tau}=-\hbar^2\Box^2K.
	\label{schro-K}
\end{equation}
This change in the measure of the path does not modify our derivation of the 
quantum correction in the semiclassical propagator. Indeed, the computation of 
the amplitude and phase of $F$ involves some products of the form $K\left[x^\mu(\tau),x^\mu(0)\right]\times{K}\left[x^\mu(0),x^\mu(\tau)\right]$ where the new terms $\sqrt{\frac{g(\tau)}{g(0)}}$ and $\sqrt{\frac{g(0)}{g(\tau)}}$ obviously cancel out. We will work in this scheme to determine the value of the constant $\alpha$ and discuss this choice in the next section.

	The computation of Eq. (\ref{schro-K}) in \cite{hartle} is carried out with a specific value of
	$\alpha$ equal to 1/4. Allowing for an arbitrary multiplicative constant, the equation of motion should take the form~:
\begin{equation}
	i\hbar{f}(\alpha)\frac{\partial{K}}{\partial\tau}=-\hbar^2\Box^2K.
\end{equation}
Following the demonstration given in the appendix of \cite{hartle}, it is straightforward to show that $f(\alpha)=4\alpha$. From this Schr\"odinger-like equation, we derive the dynamical equation driving the evolution of the {\it{complete}} propagator $G$. Applying the KG operator to $G$ leads to~:
\begin{equation}
	\hbar^2\Box^2G=-4\alpha{i}\hbar\displaystyle\int{d}\tau{ie^{\frac{i}{\hbar}\alpha\tau}\frac{\partial{K}}{\partial\tau}},
\end{equation}
where we have used Eq.~(\ref{schro-K}) to re-write the integrand as a function of the proper time. After integrating by part, one obtains~:
\begin{equation}
	\hbar^2\Box^2G=4\alpha{\hbar}\left[e^{\frac{i}{\hbar}\alpha\tau}K\right]^\infty_0-4\alpha^2G.
\end{equation}
When the proper time tends to infinity, $e^{\frac{i}{\hbar}\alpha\tau}$ is exponentially damped in the imaginary proper
time ($\tau\rightarrow{i}T$) and the upper limit in the first term of the right-hand-side of the above equation vanishes. Then, it should be noticed that the $K$ propagator becomes a Dirac distribution when the proper time tends to zero, as expected to ensure that the particle must be at its initial position at the origin of time. Consequently, the above result finally leads to the inhomogeneous KG equation, a Dirac distribution being the source term~:
\begin{equation}
	\left(\hbar^2\Box^2+4\alpha^2\right)G\left[x^\mu,x^\mu(0),\alpha\right]=\frac{\delta^4\left(x^\mu-x^\mu(0)\right)}{\sqrt{-g(0)}}.
	\label{in-kg}
\end{equation}
The source term is normalized for a particle initially at the $x^\mu(0)$ space-time position. 
To be equivalent to
quantum mechanics from a field-theory point of view, the {\it{complete}} propagator has to be a 
solution of the KG equation, leading to~:
\begin{equation}
	\alpha=\pm\frac{m}{2},
\end{equation}
where $m$ corresponds to the mass of the scalar particle. As a consequence of the quadratic HJ equation involved in relativistic mechanics, an ambiguity remains on the sign of $\alpha$. In the following, we choose  $\alpha=m/2$.

	\subsection{Dynamical equations at the semiclassical order}
	To compute the dynamical equations at the semiclassical order, both propagators (the {\it{proper-time}} and
	{\it{complete}} ones) are written as products of a classical part (the dominant path) by a quantum part (the quantum corrections)~:
\begin{eqnarray}
	\tilde{K}\left[x^\mu(\tau),x^\mu(0)\right]&=&F\left[x^\mu(\tau),x^\mu(0)\right]e^{\frac{i}{\hbar}\tilde{S}\left[x^\mu(\tau),x^\mu(0)\right]}, \\
	\tilde{G}\left[x^\mu,x^\mu(0),\alpha\right]&=&H\left[x^\mu,x^\mu(0),\alpha\right]e^{\frac{i}{\hbar}\displaystyle\int{p}_\mu{d}\bar{x}^\mu}.
\end{eqnarray}
These expressions come from the Taylor expansion around classical paths for the $\tilde{K}$ propagator and from the complete computation given in the previous section for the $\tilde{G}$ propagator. Each of the aforementioned formulas is introduced in the dynamical equations for the quantum propagator and then expanded up to the first order in reduced Planck constant. As in non-relativistic quantum mechanics, we expect the quantum equations to translate into a set of two dynamical equations~: a zeroth order equation, the HJ one, providing the underlying classical dynamics, and a first order equation involving the $F$ or the $H$ functions and providing the first order quantum corrections.

	It can also be noticed that applying the KG operator $\Box^2$ to the complete semiclassical
	propagator allows us to determine the $\alpha$ constant. Since the zeroth order term in $\hbar$ is
	$\left\{g^{\mu\nu}p_\mu{p}_\nu-4\alpha^2\right\}$, the $\alpha$ constant should be equal to
	$\pm{m}/2$ to recover the HJ equation, as given in Eq.~(\ref{H-J-momentum}).

	Performing this Taylor expansion  allows us to prove that the zeroth order leads, in both cases, 
	to the correct HJ equation, expressed using the classical action functional for the 
	$\tilde{K}$ propagator and expressed using the four-momentum for the $\tilde{G}$ propagator.
	 The first order equations then become~:
\begin{equation}
2m\frac{\partial{F}}{\partial{\tau}}=-\frac{F}{\sqrt{-g}}\partial_\mu\left[\sqrt{-g}g^{\mu\nu}\frac{\partial\tilde{S}}{\partial{x}^\nu}\right]-2g^{\mu\nu}\frac{\partial\tilde{S}}{\partial{x}^\mu}\partial_\nu{F},
\end{equation}
and
\begin{equation}
	\frac{H}{\sqrt{-g}}\partial_\mu\left[\sqrt{-g}g^{\mu\nu}p_\nu\right]+2g^{\mu\nu}p_\mu\partial_\nu{H}=0.
\end{equation}
Multiplying those equations by $F^\dag$ or $H^\dag$, depending on the considered propagator, leads to conservation equations~:
\begin{eqnarray}
	\frac{\partial\rho}{\partial{\tau}}+\frac{1}{\sqrt{-g}}\partial_\mu\left(\sqrt{-g}g^{\mu\nu}\tilde{j}_\nu\right)&=&0, \\
	\frac{1}{\sqrt{-g}}\partial_\mu\left[\sqrt{-g}g^{\mu\nu}\tilde{J}_\nu\right]&=&0.
\end{eqnarray}
On the one hand, the first of those two equations is a conservation law in 
$(4+1)$-dimensions where $\rho$ is the proper time probability density and $\tilde{j}_\mu$ is the probability current along the 4-dimensional space-time. These quantities, both at the semiclassical order, are defined by~: 
\begin{eqnarray}
	\rho&=&\left|F\right|^2, \\
	\tilde{j}_\mu&=&\frac{\left|F\right|^2}{m}\times\frac{\partial\tilde{S}}{\partial{x}^\mu}.
\end{eqnarray}
On the other hand, the second equation describes the conservation of a probability four-current in the 4-dimensional space-time. This WKB four-current is given by~:
\begin{equation}
	\tilde{J}_\mu=\left|H\right|^2p_\nu.
\end{equation}
This result is the natural extension to any curved background of the WKB four-current already obtained in the case of static and spherically symmetric curved spaces \cite{wkb}. Because the WKB four-current is proportional to the classical four-momentum, this four-vector is naturally covariant. Although the partial derivative of the action in the definition of the $\tilde{j}_\mu$ current is equal to the classical momentum, we keep this form to make clear the difference between the two current, $\tilde{j}_\mu$ and $\tilde{J}_\mu$.

We point out that the above results are still valid even if Eq. (\ref{schro-K1})
is used instead of Eq. (\ref{schro-K}). The term $\frac{\hbar^2}{3}RK$ is of
second order in Planck constant and therefore does not contribute to the
semiclassical equation of motion. This also confirms the choice of $\alpha$~: determining the value of this constant by imposing the zeroth order in $\hbar$ to be equal to the HJ equation still leads to $\pm{m}/2$ when the additional scalar curvature term is considered.

	\subsection{Domain of validity}
	The WKB propagator $\tilde{G}$ is not an exact solution of the KG equation and the domain of
	validity of the semiclassical ansatz has to be investigated. To this purpose, the
partial differential equation exactly satisfied by $\tilde{G}$ is computed and compared to the KG one. For simplicity, we
consider the pedagogical case of a one-dimensional space which allows us to re-write the WKB propagator in the following form~:
\begin{equation}
	\tilde{G}\left[x,x(0),m\right]=\frac{N}{\left|g(\tau)\right|^{1/4}\sqrt{p}}\exp{\left(\frac{i}{\hbar}\displaystyle\int{gpd\xi}\right)},
	\label{undimprop}
\end{equation}
$N$ being a normalization constant and the coordinate $\xi$ and the momentum $p$ being written in their contravariant form whereas the metric
tensor is expressed in its covariant form \footnote{This
relativistic propagator is the direct extension of the one-dimensional propagator in non-relativistic quantum mechanics
given by $\tilde{G}\left[x,x(0),\omega\right]=\frac{N}{\sqrt{p}}\exp{\left(\frac{i}{\hbar}\displaystyle\int{pdx}\right)}$.}. In such a
one-dimensional space-time, the KG operator reads~:
\begin{equation}
	\Box^2=\frac{1}{\sqrt{\left|g\right|}}\frac{\partial}{\partial\xi}\left[\sqrt{\left|g\right|}g^{-1}\frac{\partial}{\partial\xi}\right].
\end{equation}

The inverse of $g$ appears as the metric tensor is involved in its contravariant form in the KG equation. Applying the above operator on the WKB propagator leads to~:
\begin{equation}
	\hbar^2\Box^2\tilde{G}=\left\{-gp^2+\hbar^2\Delta_{p}+\hbar^2\Delta_{gp}+\hbar^2\Delta_{g}\right\}\tilde{G},
\end{equation}
with
\begin{eqnarray}
	\Delta_{p}&=&\frac{3}{4gp^2}\left(\frac{\partial{p}}{\partial\xi}\right)^2-\frac{1}{2gp}\frac{\partial^2{p}}{\partial\xi^2}, \\
	\Delta_{gp}&=&\frac{1}{2g^2p}\left(\frac{\partial{p}}{\partial\xi}\right)\left(\frac{\partial{g}}{\partial\xi}\right)-\frac{1}{4g\left|g\right|p}\left(\frac{\partial{p}}{\partial\xi}\right)\left(\frac{\partial{\left|g\right|}}{\partial\xi}\right), \\
	\Delta_{g}&=&\frac{1}{4g^2\left|g\right|}\left(\frac{\partial{\left|g\right|}}{\partial\xi}\right)\left(\frac{\partial{g}}{\partial\xi}\right)+\frac{1}{16g\left|g\right|^{3/2}}\left(\frac{\partial{\left|g\right|}}{\partial\xi}\right)^2.
\end{eqnarray}
As for a one-dimensional system $gp^2=m^2$, the WKB propagator is an exact
solution of~:
\begin{equation}
	\left[\hbar^2\Box^2+m^2\right]\tilde{G}=\hbar^2\left\{\Delta_{p}+\Delta_{gp}+\Delta_{g}\right\}\tilde{G}.
	\label{wkbexact}
\end{equation}
For the semiclassical ansatz to be a good approximation, the right-hand side of the above equation should be very small as compared to
$m^2$. This also proves, as in non-relativistic quantum
mechanics, that the approximation necessarily breaks down at focal or turning points, where the momentum vanishes, as the right-hand side diverges at those precise points.

	The $\Delta_i$ functions involved in equation (\ref{wkbexact}) can be 
	interpreted as the error inherent to the semiclassical
propagator. This error is due to the variation along the paths of the momentum, the $\Delta_p$ function, as well as the variation of the metric tensor,
the $\Delta_g$ function. This also induces a third part to the error, the $\Delta_{gp}$ function, which corresponds to the
cross-variation of the $g$ and $p$ quantities. If the background is flat, only
the momentum variation contributes to the error and the validity condition
reads~:
\begin{equation}
	\left|m^2\right|\gg\left|\frac{3}{4p^2}\left(\frac{\partial{p}}{\partial\xi}\right)^2-\frac{1}{2p}\frac{\partial^2{p}}{\partial\xi^2}\right|,
	\label{validity}
\end{equation}
which is in total agreement with the condition of validity derived for the propagation of scalar fields in Schwarzschild-like space-times \cite{wkb}, as it can be seen
comparing the right-hand side of Eq. (\ref{validity}) and the right-hand side of
Eq. (3) in Ref. \cite{wkb}. It should be reminded that in
\cite{wkb}, the propagation is treated as in flat spaces using the tortoise coordinates system and taking into account a potential-like term mimicking the effect of the
gravitational and centrifugal potential. This explains the difference in the
left-hand side of the validity conditions. Assuming now a
propagation in a $(1+1)$-dimensional flat space-time and working in the temporal frequency domain, {\it{i.e.}}
$\tilde{G}\equiv{e}^{i\omega{t}}\tilde{g}(x,x(0))$, the spatial part of the propagator can still be written in the form of Eq.
(\ref{undimprop}) where $\xi$ and $p$ stands for the spatial coordinate and momentum. With such a separation of variables, the condition
of validity for the WKB propagator reads as in Eq. (\ref{validity}), replacing $\left|m^2\right|$ by
$\left|m^2-\omega^2\right|=\left|p^2\right|$. This latest condition is in agreement with the one proposed in Ref. \cite{wkb}.

\section{Discussion and comparison with other results}
\label{sec3}
	As already mentioned earlier in this article, the definition and computation of the 
	propagator for scalar particles in curved space-times is ambiguous. The above
	 construction of the WKB propagator is mainly motivated by an analogy between 
	 general relativistic mechanics and Newtonian mechanics, though some
	alternative ways are possible. Therefore, our results have to be
	compared with different approaches. For this purpose, it is firstly 
shown that the WKB propagator derived in this work leads to the right Feynman 
propagator in the limit of flat spaces. Then, we clarify the analogy adopted 
between the {\it{complete, general relativistic}} propagator and the propagator 
for stationary systems in non-relativistic quantum mechanics. Finally, the 
results obtained are compared with the ones used in \cite{dewitt,dewitt1,hartle} focusing our discussion on the additional scalar curvature term involved in the equation of motion (see Eq. (\ref{schro-K1})) and on the choice of the weight when passing from the {\it{proper time}} propagator to the {\it{complete}} one.

	\subsection{Changing the action functionnal}
	In the semiclassical approximation, the propagator can be cast into a 
form which does not depend on the choice of the
action functional \footnote{Semiclassical schemes do not require priors for 
the form of the action functional, except
	that it satisfies classical mechanics.}~:
\begin{equation}
	\tilde{G}\left[x^\mu,x^\mu(0),\alpha\right]=\frac{i\left|\det_{\mu\nu}{\left[\frac{\partial^2\tilde{S}}{\partial{x}^\mu\partial{x}^\nu(0)}\right]}\right|^{1/2}}{\sqrt{\left(2i\pi\hbar\right)^3\left.\frac{\partial^2\tilde{S}}{\partial\tau^2}\right|_{\bar{\tau}}\sqrt{g(\bar\tau)g(0)}}}\exp{\left(\frac{i}{\hbar}\displaystyle\int^{x^\mu}_{x^\mu(0)}g_{\mu\nu}p^\nu{d}\bar{x}^\mu-iq\frac{\pi}{2}\right)}.
\end{equation}
It is therefore needed to investigate the influence of different action functional choices, still compatible with general relativity, on the WKB
propagator. It is mentioned in \cite{hartle} that choosing an action functional quadratic in four-velocity allows us to recover the
usual relativistic quantum mechanics of scalar particles. We re-address 
hereafter this problem by studying the WKB propagator for
free, scalar particles evolving in a flat space-time under two different assumptions for the
action functional~: quadratic or linear in four-velocity.

	If the usual definition for the action functional is chosen instead of
	the quadratic one, the canonical four-momentum and the HJ equation
	read~:
\begin{equation}
	p_\mu=m\dot{x}_\mu,
\end{equation}
and
\begin{equation}
	g_{\mu\nu}p^\mu{p}^\nu=m^2.
\end{equation}
Computing the partial derivatives of $\tilde{S}$ according to the proper time
leads to~:
\begin{equation}
	\frac{\partial\tilde{S}}{\partial\tau}=0.
\end{equation}
Consequently, no conjugate variable can be associated with the proper time
derivative of the classical action and the HJ equation is reduced to~:
\begin{equation}
	g^{\mu\nu}\left(\frac{\partial\tilde{S}}{\partial{x}^\mu}\right)\left(\frac{\partial\tilde{S}}{\partial{x}^\nu}\right)=m^2.
\end{equation}
Such a behavior clearly disfavors an action functional linear in four-velocity, though it does not rule out our definition of the
propagator, as it simply consists in taking $\alpha$ equal to zero when
performing the final proper time integration leading to the complete propagator given by Eq. (\ref{complete-prop}).

	For a freely-moving particle in a flat background, the semiclassical approximation is exact (derivatives of the action higher
than the second order ones vanish) and the four-velocity becomes a constant~:
\begin{equation}
	\dot{x}^\mu(\tau)=\frac{\Delta{x}^\mu}{\tau},
\end{equation}
where $\Delta{x}^\mu=x^\mu(\tau)-x^\mu(0)$ stands for the four-interval between
the final and initial events. With this four-velocity, the four-momentum,
whatever the action functional considered, is also constant along the
trajectory and the classical action is given by~:
\begin{eqnarray}
	\tilde{S}_l\left[x^\mu(\tau),x^\mu(0)\right]&=&m\sqrt{\eta_{\mu\nu}\Delta{x}^\mu\Delta{x}^\nu}, \\
	\tilde{S}_q\left[x^\mu(\tau),x^\mu(0)\right]&=&\frac{m}{2\tau}\eta_{\mu\nu}\Delta{x}^\mu\Delta{x}^\nu,
\end{eqnarray}
where the subscript $l(q)$ means the {\it{linear(quadratic)}} choice \footnote{We can notice the direct analogy between the quadratic action functional and classical mechanics for a free particle, whose action is $\tilde{S}_c(\vec{x}(t),\vec{x}(0))=m\left|\vec{x}(t)-\vec{x}(0)\right|^2/{2(t-t_0)}$.}.  From these actions, we get the following quantum correction
\begin{eqnarray}
	\left|F_l\right|^2\left[x^\mu(\tau),x^\mu(0)\right]&=&\left(\frac{m}{2\pi\hbar\Delta_\tau}\right)^4\times\left(1+\frac{\det_{\mu\nu}{\left[\Delta{x}^\mu\Delta{x}^\nu\right]}}{\Delta_\tau^8}\right), \\
	\left|F_q\right|^2\left[x^\mu(\tau),x^\mu(0)\right]&=&\left(\frac{m}{2\pi\hbar\tau}\right)^4,
\end{eqnarray}
where $\Delta_{\tau}=\sqrt{\eta_{\mu\nu}\Delta{x}^\mu\Delta{x}^\nu}$ stands for the scalar, Minkowskian interval. Because of the $\tau^{-2}$
component in the $F$ functional, an unidimensional convergence parameter $\varepsilon$ has to be introduced when performing the integration over
the proper time, insuring the convergence for vanishing proper time paths~:
\begin{equation}
	G\equiv{i}\displaystyle\int^\infty_0\exp{\left(\frac{i}{\hbar}\alpha\tau-\frac{\hbar\varepsilon}{\alpha\tau}\right)}K[x^\mu(\tau),x^\mu(0)]d\tau.
\end{equation}
We re-introduce the $\alpha$ notation in the above equation as it changes from one choice to the other for the action.

	In the case of a functional quadratic in four-velocity, the
	aforementioned integral writes~:
\begin{equation}
	G=\frac{-im^2}{4\pi^2\hbar^2}\displaystyle\int^\infty_0\frac{e^{\frac{im\Delta_\tau^2}{2\hbar\tau}}}{\tau^2}\times{e}^{\frac{im}{2\hbar}\tau-\frac{\hbar{m}\varepsilon}{2\tau}}d\tau.
\end{equation}
Changing the time variable as $u=\tau/2m$, the {\it{complete}} propagator in flat spaces is given by the same integral
as the one computed in \cite{hartle} (see equations (2.19) and (3.2)),
where we explicitly keep the reduced Planck constant~:
\begin{equation}
	G=\frac{-2im}{\left(4\pi\hbar\right)^2}\displaystyle\int^\infty_0\frac{e^{\frac{i\Delta_\tau^2}{4\hbar{u}}}}{u^2}\times{e}^{\frac{im^2}{\hbar}u-\frac{\hbar\varepsilon}{u}}du,
\end{equation}
where the convergence parameter has been rescaled by $\varepsilon\rightarrow4\varepsilon$. This rescaling of the convergence
parameter is only a matter of convention. As shown in \cite{hartle} (see equation (3.3)), the above integral on the $u$ variable leads to the right
Feynman propagator
\begin{equation}
	G\left[x^\mu,x^\mu(0),m\right]=\left(\frac{-2im}{4\pi^2}\right)\frac{1}{\Delta^2_\tau-i\varepsilon},
\end{equation}
where we set the reduced Planck constant equal to unity for a direct comparison with the propagator derived in \cite{hartle}. Because we
choose the $(+---)$ convention, the convergence parameter enters with the $(-)$ sign instead of the $(+)$ one in
\cite{hartle} where the $(-+++)$ convention is used. The normalization constant of our propagator is also different from the one
in Eq. (3.3) of \cite{hartle} because we adopt a different weight than \cite{hartle} during the computation of the $G$
propagator. 

	The computation of the {\it{complete}} propagator with an action functional linear in four-velocity is useless.
The {\it{quadratic-defined}} $K$ propagator leads to the right Feynman propagator and is very different form the
{\it{linear-defined}} one. Choosing an action functional linear in four-velocity
 cannot lead to the right Feynman propagator in
flat spaces whereas a quadratic action can.

	\subsection{Propagator in non-relativistic quantum mechanics}
	The WKB propagator describing the dynamics of a scalar particle in any
 curved background is very reminiscent of the one
	describing a time-independent, non-relativistic quantum system at the semiclassical order. This is a direct consequence of the
	way we define the aforementioned quantity. Calling $\vec{x}$ the three-dimensional spatial vector and $t$ the time, the non-relativistic WKB propagator $\tilde{K}$ and its Fourier transform in the frequency domain $\tilde{G}$, 
	are defined by~:
\begin{eqnarray}
	\tilde{K}(\vec{x}(t),\vec{x}(0))&=&F(\vec{x}(t),\vec{x}(0))e^{\frac{i}{\hbar}\tilde{S}(\vec{x},\vec{x}(0);t,0)}; \\
	\tilde{G}(\vec{x},\vec{x}(0),\omega)&=&\displaystyle\int^\infty_0\tilde{K}(\vec{x},\vec{x}(0);t,0)e^{\frac{i}{\hbar}\omega{t}}dt.
\end{eqnarray}
Using the {\it{fixed energies}} propagator $\tilde{G}$ makes sense only in the 
case of time-independent, {\it{i.e.}} stationary, systems
where the energy is the relevant variable describing particles. From these definition, we can underline a direct analogy between the
aforementioned propagator and the {\it{general relativistic}} one. This analogy, depicted on table \ref{analogy}, shows that the proper
time and the four-coordinates are respectively analogous to the usual time and to the three-dimensional spatial vector in non-relativistic
quantum mechanics. This also translates into an analogy between conjugate variables, the four-momentum being in correspondence with the
three-dimensional momentum and the constant $\alpha=f(m)$ is equivalent to the energy $\omega$ used in non-relativistic quantum mechanics.
Such a correspondence obviously justifies the transformation from a
{\it{fixed-proper-time}} propagator to a {\it{fixed mass}} propagator,
already enlightened in the introduction.
\begin{table}[ht]
	\begin{center}
		\begin{tabular}{c|c} \hline\hline
			non-relativistic & general relativistic \\
			quantum mechanics & quantum mechanics \\ \hline\hline
			3-space vector & 4-space-time coordinates \\
			$\vec{x}(t)$ & $x^\mu(\tau)$ \\ \hline
			usual time & proper time \\
			$t$ & $\tau$ \\\hline
			3-momentum & 4-momentum \\
			$\vec{p}=\vec{\nabla}\tilde{S}$ & $p_\mu=\frac{\partial\tilde{S}}{\partial{x}^\mu}$ \\\hline
			particle's energy & particle's mass \\
			$\omega$ & $m$ \\\hline
			classical action & relativistic action \\
			$S=\int\frac{mV^2}{2}dt$ & $S=\int\frac{m}{2}{g}_{\mu\nu}\dot{x}^\mu\dot{x}^\nu{d}\tau$ \\\hline
			{\it{fixed-time}} propagator & {\it{fixed-proper-time}} propagator \\
			$K(\vec{x}(t),\vec{x}(0))$ & $K[x^\mu(\tau),x^\mu(0)]$ \\\hline
			{\it{fixed-energy}} propagator & {\it{complete}} propagator \\
			$G(\vec{x},\vec{x}(0),\omega)$ & $G[x^\mu,x^\mu(0),m]$ \\ \hline
			probability density and current & probability four-current \\
			$\rho=\left|F\right|^2$ and $\vec{\jmath}=\frac{\left|F\right|^2}{m}\vec{p}$ & $J_\mu=\left|H\right|^2p_\mu$ \\ \hline\hline
		\end{tabular}
		\caption{Analogy between non-relativistic quantum mechanics and general relativistic quantum mechanics. This analogy 
		shows the reminiscence between the computation of the semiclassical propagator for any curved background and the computation of the semiclassical propagator, at fixed energies, for a non-relativistic, stationary quantum system.}
		\label{analogy}
	\end{center}
\end{table}

	The same kind of analogy is drawn out from the semiclassical equation
	of motion which, for non-relativistic quantum mechanics, takes also the
	form of a conservation equation~:
\begin{equation}
	\frac{\partial\left|F\right|^2}{\partial{t}}+\vec{\nabla}\cdot\left(\frac{\left|F\right|^2}{m}\vec{p}\right)=0.
\end{equation}
In a non-relativistic framework, the square modulus of $F$ is interpreted as the probability density and the product of this square modulus
times the classical velocity as the probability current. Interpreting, in the general relativistic case, the product of the square modulus
of $F$ times the four-momentum is a natural generalization of the aforementioned non-relativistic interpretation, 
just as the generalization of the Schr\"odinger equation to the KG one. Nevertheless, we can notice 
the direct analogy between the non-relativistic quantum propagator with the {\it{proper-time}} 
propagator~: in the $(4+1)-$description, the square modulus of $F$ and its product with the four-velocity 
still stand for the probability density and for the probability current respectively, 
$\tau$ playing the role of the usual time and the space-time four coordinate the role of the 3-dimensional space.

	This behavior, as well as the analogy mentioned in this section, underline what was argued in 
the introduction~: in curved spaces, the {\it{proper-time}} propagator is analogous to the 
{\it{fixed-time}} propagator in non-relativistic quantum mechanics, whereas the {\it{complete}} 
propagator is equivalent to the {\it{fixed-energy}} propagator used to describe stationary quantum 
systems. Such an analogy lies in our description of classical mechanics~: the motion of a particle is
 parametrized by an affine parameter which is the usual time in non-relativistic mechanics and the 
 proper time $\tau$ in general relativistic mechanics. The conjugate variable of this affine parameter, 
 energy in non-relativistic mechanics and mass in general relativistic mechanics, are equivalent, especially when computing the propagator.

\subsection{Other results for the propagation in curved spaces}

\begin{flushleft}
{\underline{Propagator in the small proper time limit and choice of the proper
time weight~:}}
\end{flushleft}
	The {\it{complete}} propagator for scalar particles evolving in a curved background has 
been studied in its fully quantum form to determine the Hawking's temperature \cite{hartle}. The 
authors of Ref. \cite{hartle} also mentioned that for small proper time motions, the propagator is 
well approximated by its WKB expansion given by \cite{dewitt}~:
\begin{equation}
	\tilde{G}\left[x^\mu,x^\mu(0),m\right]=\frac{-i}{\left(4\pi\right)^2}\left|\det_{\mu\nu}{\left[\frac{\partial^2\tilde{S}}{\partial{x}^\mu\partial{x}^\mu(0)}\right]}\right|^{1/2}\exp{\left(\frac{i}{\hbar}\tilde{S}-\frac{i}{\hbar}m^2\bar{\tau}+\Omega\left[x^\mu,x^\mu(0)\right]\right)},
\end{equation}
where $\Omega$ is a function determined by the semiclassical equation of motion. The results
 depicted above are different from the one we derived, the differences arising in the quantum part 
(the $F$ functional) as well as in the classical one (the $-im^2\bar{\tau}$ in the exponential). 

The difference in the $F$ functional corresponds to the appearance of the $(g(\tau)g(0))^{-1/2}$
 prefactor in our results. However it should be noticed that the above WKB propagator stands for
  small proper times for which the metric tensor is equal to the Minkowski one at the zeroth order. 
If our propagator is used to study a small proper time motion, the aforementioned assumption is 
valid and the $(g(\tau)g(0))^{-1/2}$ prefactor becomes equal to one. Nevertheless, when one 
wants to approximate, at the WKB order, the quantum propagation for arbitrary
proper times, our 
prefactor should be kept in order to insure the propagator to be a scalar. 

The second point is more complicated as it relies on different assumptions for the classical 
action and the definition of the propagator. In \cite{hartle,dewitt}, the {\it{complete}} 
propagator is defined as~:
\begin{equation}
	G\equiv{i}\displaystyle\int{K}\exp{\left(-\frac{i}{\hbar}m^2\tau\right)},
\end{equation}
with an action functional of the form~:
\begin{equation}
	\tilde{S}\equiv\frac{1}{4}\displaystyle\int{g_{\mu\nu}}\dot{x}^\mu\dot{x}^\nu{d\tau}.
\end{equation}
With this definition, the propagator is still a solution of the inhomogeneous KG equation (see Eq. (\ref{in-kg})). 
First of all, the multiplicative factor in the action differs from our one and
therefore the first derivative of the classical action according to the proper
time becomes $-1/4$. Applying our method to convert {\it{fixed-proper-time}}
paths into {\it{fixed-mass}} paths with the aforementioned action would lead to~:
\begin{equation}
	\mathcal{G}\equiv{i}\displaystyle\int{K}\exp{\left(\frac{i}{4\hbar}\tau\right)}.
\end{equation}
Clearly, this propagator $\mathcal{G}$ is not a solution of the KG equation for a field with a mass $m$ but of a KG equation with a mass equal
to one half. This justifies the $\exp{(-im^2\tau)}$ in the propagator of \cite{hartle,dewitt}. 
Nevertheless, as long as the multiplicative factor in the action is a real
number, independent of the mass, defining the {\it{complete}} propagator by~:
\begin{equation}
	G\equiv{i}\displaystyle\int{K}\exp{\left(\frac{i}{\hbar}\frac{m^2}{4\beta}\tau\right)}d\tau,
\end{equation}
where $\beta$ is the multiplicative factor in the action functional ($\beta=1/4$ in \cite{hartle}), leads to a propagator solution of the
inhomogeneous KG equation. It therefore seems that defining the propagator following \cite{hartle} is more ambiguous.

	The semiclassical expansion is based on the stationary phase approximation which imposes 
to weight the {\it{proper-time}} propagator with the conjugate variable
associated with the proper time~:
\begin{equation}
	G\equiv{i}\displaystyle\int{K}\exp{\left(-\frac{i}{\hbar}\frac{\partial\tilde{S}}{\partial\tau}\tau\right)}.
\end{equation}
With the above definition, applying the stationary phase hypothesis is straightforward and we argue that our propagator should be preferred to
the one proposed in \cite{hartle}, at least for semiclassical studies, as it clearly bridges the gap between classical and quantum dynamics. In
addition, this definition, very reminiscent of the semiclassical approximation in non-relativistic quantum mechanics, allows us to unambiguously
determine, up to its sign, the multiplicative factor entering the action functional by requiring 
that the path integral approach is equivalent
to the quantum field approach. The alternative definition proposed in this work avoids the ambiguity
of \cite{hartle} by
imposing an additional constrain inspired by non-relativistic quantum mechanics~: the weight to 
convert a {\it{fixed-proper-time}} propagator $K$ into
{\it{fixed-mass}} propagator $G$ is a function of the conjugate variable of the 
proper time $\tau$.

Finally, it should be noticed that the definition of the action functional in \cite{hartle} is not
 homogeneous to $\hbar$ whereas our definition is. A possible {\it experimental}
approach to discriminate between the two proposals will be presented in the 
next section.

\begin{flushleft}
{\underline{Effect of the scalar curvature term~:}}
\end{flushleft}
	The appearance of the Ricci scalar as a potential-like term in the equation of motion 
for the {\it{proper time}} propagator is a direct consequence of the path integral formulation of 
the quantum theory. Although this curvature term does not modify the
semiclassical dynamics, the definition of the multiplicative constant in the
action functional has been determined in a scheme where the $R$ term is not
present. Relaxing this constrain, Eq. (\ref{in-kg}) is given by~:
\begin{equation}
	\left[\hbar^2\Box^2\hbar^2-\lambda{R}+4\alpha^2\right]G=\frac{\delta^4\left(x^\mu-x^\mu(0)\right)}{\sqrt{-g(0)}},
	\label{in-kg1}
\end{equation}
where $\lambda$ is a constant which depends on the way the infinitesimal propagator is defined (see appendix of \cite{hartle}). Claiming that the $\alpha$ constant is deduced by imposing $4\alpha^2-\hbar^2\lambda{R}=m^2$ is spurious as it would mean that the mass of scalar particles depends on the space-time where they 
evolve. In addition, it is shown in \cite{dewitt1} that the curvature term arises because of 
{\it{curvature-induced}} non-linearities in the dynamics of the propagator. The dynamical origin 
of $R$ in the equation of motion prevents us from determining $\alpha$ thanks to 
$4\alpha^2-\hbar^2\lambda{R}=m^2$. Nevertheless, we argue that $\alpha$ is still equal to
 $\pm{m}/2$. Since the dynamical equation (\ref{schro-K1}) is derived using a semiclassical
 structure for the propagator (see Eq. (7.1) in \cite{dewitt1}), it is more
 meaningful to determine $\alpha$ thanks to semiclassical arguments~: this prefactor is computed by forcing the zeroth order part of the $\hbar$ expansion of the equation of motion to be the HJ one.

	Keeping $\alpha=\pm{m}/2$ leads to an inhomogeneous KG equation where a
	potential-like term arises~:
\begin{equation}
	\left[\hbar^2\Box^2-\hbar^2\lambda{R}+m^2\right]G=\frac{\delta^4\left(x^\mu-x^\mu(0)\right)}{\sqrt{-g(0)}}.
	\label{in-kg2}
\end{equation}
In a field theory approach, the curvature term can consequently be interpreted as a coupling 
between curvature and fields. Indeed, a scalar field theory described by the
action~:
\begin{equation}
	S\left[\Phi(x^\mu)\right]=\displaystyle\int\sqrt{-g}d^4x\left[g^{\mu\nu}\frac{\partial\Phi}{\partial{x}^\mu}\frac{\partial\Phi}{\partial{x}^\nu}-m^2\Phi^2+\lambda{R}\Phi^2\right],
\end{equation}
leads to Euler-Lagrange equations given by Eq. (\ref{in-kg2}) without the source term. It is interesting to point out 
this coincidence between the path integral formulation of quantum mechanics and the field 
theory approach with some couplings between scalars and gravity. It is however a naive 
conclusion to deduce that a field theory viewpoint for quantum mechanics in curved spaces has
to contain couplings to gravity because the path integral formulation leads to a similar term. On 
the one hand, unlike the aforementioned action for fields, the action functional for point particles
 chosen here corresponds to a minimal coupling to gravity. On the other hand, the Euler-Lagrange 
 equation derived from the above action functional stands for a classical description of fields and 
 a careful comparison between both formulations of quantum mechanics should be done at the level of
  propagators.

\section{A possible dedicated experiment}
\label{interference}
	
Two different candidate propagators are therefore considered~:
\begin{eqnarray}
	\tilde{G}\left[x^\mu,x^\mu(0),\alpha\right]&=&\frac{i\sqrt{\mathcal{H}\left[x^\mu(\bar{\tau}),x^\mu(0)\right]}}{\sqrt{\left(2i\pi\hbar\right)^3\sqrt{g(\bar{\tau})g(0)}}}\exp{\left(\frac{i}{\hbar}\displaystyle\int^{x^\mu}_{x^\mu(0)}g_{\mu\nu}p^\nu{d}\bar{x}^\mu-iq\frac{\pi}{2}\right)}, \\
	\tilde{G}\left[x^\mu,x^\mu(0),m\right]&=&\frac{-i}{\left(4\pi\right)^2}\left|\det_{\mu\nu}{\left[\frac{\partial^2\tilde{S}}{\partial{x}^\mu\partial{x}^\mu(0)}\right]}\right|^{1/2}\exp{\left(\frac{i}{\hbar}\tilde{S}-\frac{i}{\hbar}m^2\bar{\tau}+\Omega\left[x^\mu,x^\mu(0)\right]\right)}.
\end{eqnarray}
The keypoint to distinguish between them is to have access to the phase in a curved
background.
The two considered proposals differ by the weights associated to each path at fixed proper time
$\tau$~:  $\left[e^{im\tau/2}\right]$ in the first case and
$\left[e^{im^2\tau}\right]$ in the second case. We suggest to measure this
difference, which enters directly the phase of the propagator, by considering
quantum interferences due to the gravitational field, {\it i.e.}
curvature-induced.

Let us consider an interferometer made of two arms $(1)=a\to{b}\to{d}$ and
$(2)=a\to{c}\to{d}$, orbiting a spherical source of gravitation. The background
metric is given by~:
\begin{equation}
	ds^2=B(r)dt^2-A(r)dr^2-r^2d\theta^2-r^2\sin^2(\theta)d\varphi^2.
\end{equation}
In the equatorial plane $\theta=\pi/2$, the arms are in the following configuration~: 
$a\to{b}$ and $c\to{d}$ (the {\it radial} arms) run from $r_a$ to $r_d$ whereas 
$a\to{c}$ and $b\to{d}$ (the {\it orbital} arms) are according to equatorial orbits 
respectively at $r_a$ and $r_d$. The orbital arms are assumed to be short enough so
that the classical orbits can be approximated by portions of circles. In this case, the
classical trajectories at fixed radius are determined by~:
\begin{eqnarray}
	0&=&\frac{J^2}{r^2}-\frac{1}{B(R)}+E(r),\\
	E(r)&=&B^{-1}(r)\left(1-\frac{rB'(r)}{2B(r)}\right),\\
	J^2(r)&=&\frac{B'(r)r^3}{2B^2(r)},\\
	\frac{d\varphi}{dt}&=&\sqrt{\frac{B'(r)}{2r}},\\
	\frac{d\tau}{dt}&=&\sqrt{B(r)-\frac{1}{2}rB'(r)}.
\end{eqnarray}
If a coherent flux of massive particles is injected within the interferometer,
interferences should occur in $d$ as the complete propagator is given by the 
sum over the two possible classical paths~:
\begin{equation}
	K(a\to{d})=K(a\to{b}\to{d})+K(a\to{c}\to{d}).
\end{equation}
To compute the interference pattern between particles which went through the arm (1)
and particles which went through the arm (2), we follow the semiclassical
approximation. The phase of the propagators is then given by the classical action along
the classical path together with the turning points contributions. The squared modulus
of the propagator is then given by~:
\begin{eqnarray}
	\left|K(a\to{d})\right|^2&\propto&\left|F_1+F_2e^{i(\phi_2-\phi_1)}\right|^2, \\
	&\propto&\left|F_1\right|^2+\left|F_1\right|^2+2F_1F_2\cos{\underbrace{\left(\phi_2-\phi_1\right)}_{\Delta\phi}}
\end{eqnarray}
where $\phi$ are the phases and $F_i$ are real-valued functions whose squared moduli
are proportional to the probability to go through the arm $i$. Each arm contains two
turning points, in $a$ and $b$ for the arm (1), and in $c$ and $d$ for the arm (2). The
turning points do not contribute do $\Delta\phi$. The change of the action is 
identical along the two radial arms ($a\to{b}$ and $c\to{d}$) and no phase 
shift is to be expected from this part of the paths. But, the curvature
being different, the phase difference in the orbital paths is not the same, leading to
a phase shift which depends on the radii $r_a$ and $r_d$. In the experimental setup
considered here, the azimuthal angle variation  $\delta\varphi$ is the same for the
paths  $b\to{d}$ and $a\to{c}$. The action variation can therefore be computed at fixed
$r$ as a function of $\delta\varphi$.

For the propagator weighted with $\left[e^{im\tau/2}\right]$, taking into account the
characteristics of classical orbital trajectories, this leads to~:
\begin{eqnarray}
	\phi_{m/2}&=&\displaystyle\int^t_0mB(r)\frac{dt}{d\tau}dt\\
	&=&m\sqrt{\frac{2r}{E(r)B'(r)}}\delta\varphi.
\end{eqnarray}
For the propagator weighted with $\left[e^{im^2\tau}\right]$, this leads to~:
\begin{eqnarray}
	\phi_{m^2}&=&\displaystyle\int^{\bar\tau}_0\left[\frac{B(r)}{4}\left(\frac{dt}{d\tau}\right)^2-m^2\right]d\tau\\
	&=&\left[\frac{1}{4}-m^2E(r)B(r)\right]\sqrt{\frac{2r}{E(r)B'(r)}}\delta\phi.
\end{eqnarray}
It can be seen that, depending on the chosen weight, the quantum phase along different
orbits varies and could therefore be probed by an interference experiment. In the
considered example, this would give~:
\begin{eqnarray}
	\Delta\phi_{m/2}&=&m\left[\sqrt{\frac{2r_a}{E(r_a)B'(r_a)}}-\sqrt{\frac{2r_d}{E(r_d)B'(r_d)}}\right]\delta\varphi,\\
	\Delta\phi_{m^2}&=&\left[\left(\frac{1}{4}-m^2E(r_a)B(r_a)\right)\sqrt{\frac{2r_a}{E(r_a)B'(r_a)}}\right.\\
	&&\left.-\left(\frac{1}{4}-m^2E(r_d)B(r_d)\right)\sqrt{\frac{2r_d}{E(r_d)B'(r_d)}}\right]\delta\varphi. \nonumber
\end{eqnarray}
Whatever the chosen approach to build the propagator, the interference pattern will
depend upon 3 parameters linked with the experimental setup ($r_a,r_d$ and
$\delta\varphi$), upon the gravitational parameters entering the functions $B(r)$ and
$E(r)$, and upon the mass $m$ of the particle. Clearly, this experiment could
allow us to distinguish between the proposals.

It is worth noticing that this kind of experiment should also be a powerful test
of alternative theories of gravitation. If, {\it e.g.}, the metric function
is given by the Eddington-Robertson expansion, 
\begin{eqnarray}
	B(r)=1-\alpha\frac{2GM}{r}+\frac{1}{2}(\beta-\alpha\gamma)\left(\frac{2GM}{r}\right)^2+\cdots
\end{eqnarray}
the interference pattern will be very sensitive to the chosen values for the
parameters $\alpha, \beta, \gamma$, etc.\\

More importantly, a possible discrimination means (without requiring an
explicit experiment) could be found by considering the Newtonian limit.
Explicitly, the {\it classical} relativistic dynamics should allow us, in the
appropriate limit, to recover the {\it classical} Newtonian dynamics up to a full
derivative. The semiclassical propagator is therefore expected to take the
following form~:
\begin{equation}
	\tilde{K}\to{F}[x(t),x(0)]\exp{\left(\frac{i}{\hbar}\displaystyle\int^t_0{L_{N}(x,\dot{x})}dt\right)},
\end{equation}
where $L_{N}=\frac{1}{2}mV^2-m\Psi$  is the Newtonian Lagrangian with $\Psi$ the gravitational potential. The phase of the propagator weighted with $\left[e^{im\tau/2}\right]$
will be~:
\begin{eqnarray}
	\displaystyle\int^{x^\mu}_{x^\mu(0)}g_{\mu\nu}p^\mu{d}\bar{x}^\nu&\simeq&-m\displaystyle\int^t_0\left(\frac{1}{2}V^2-\Psi-1\right)dt \\
	&\simeq&-\displaystyle\int^t_0\left(\underbrace{{L_{N}(x,\dot{x})}-m}_{\tilde{L}_{N}}\right)dt.
\end{eqnarray}
Up to a constant $m$, the Lagrangian assumes the correct limit. This can be
straightforwardly understood if one considers the associated Hamiltonian 
$\tilde{H}_{N}=m+\frac{1}{2}mV^2+m\Psi$ which allows us to interpret  the $m$
term as the mass energy. Several conclusions can be drawn out from those remarks.
Basically, the $\left[e^{im\tau/2}\right]$ propagator leads (up to the mass term) to
the correct Lagrangian in the Newtonian limit. This is not true anymore with 
the proposal in \cite{dewitt} which leads to the limit~: 
$\tilde{L}_{N}=(m^2+3/4)(\frac{V^2}{2}-\Psi)-(m^2+1/4)$. Of course, both
Lagrangians lead to the correct Euler-Lagrange equations in the gravitational
field but the prefactors do influence the quantum phases of propagators.
Explicitly, neutron interferences in the Earth gravitational field have been
measured (see {\it e.g.} \cite{neutrons}). The resulting interference pattern 
can be explained by a phase difference  $\Delta\phi=m(\Psi(r_d)-\Psi(r_a))$, as
it can obviously be expected from the usual Lagrangian $\frac{1}{2}mV^2-m\Psi$ where the
kinetic term does not contribute because the speed is the same in both arms of
the interferometer. Of course, the mass term of the relativistic propagator does
not contribute either just because particles spend the same time in each arm.
This phase difference can be compared with the Newtonian limit of both
propagators. The one weighted $\left[e^{im\tau/2}\right]$ leads to the very same result 
whereas the one weighted $\left[e^{im^2\tau}\right]$ leads to
$(m^2+3/4)(\Psi(r_d)-\Psi(r_a))$. This disfavors propagators as defined in \cite{dewitt}.

Furthermore, the relativistic mass term could be observed if one considers an experimental set-up
where particles spend a different amount of time in each arm. Let us consider an
interferometer with two orbital arms seen at the same angular aperture  $\delta\varphi$
from the center of the source of gravitation. Clearly, the time spend in each arm will not
be the same. A Newtonian computation leads to~:
\begin{equation}
	\Delta\phi_N=\phi_N(r_d)-\phi_N(r_a),
\end{equation}
with
\begin{equation}
	\phi_N(r)=-\frac{mGM}{2r}\sqrt{\frac{r^3}{GM}}\delta\varphi,
\end{equation}
whereas a relativistic computation (with weight $\left[e^{im\tau/2}\right]$) gives in the
Newtonian limit~:
\begin{equation}
	\tilde\phi_N(r)=m\left(1-\frac{GM}{2r}\right)\sqrt{\frac{r^3}{GM}}\delta\varphi.
\end{equation}
This opens exciting perspectives to experimentally probe the quantum phases in the action
with interferometers.

\section{Conclusion}
	Motivated by a simple and analytical, though approximated, formulation of quantum propagation in gravitational fields, a complete
derivation of the semiclassical or WKB propagator for scalar particles evolving in any curved space is given in this work. For
this purpose, we have taken advantage of semiclassical results already known in non-relativistic quantum mechanics as well as for quantum
propagation in static and spherically, {\it{i.e.}} Schwarzschild-like, backgrounds. The analogy between proper time/four-coordinates,
$(\tau,x^\mu$), in general relativity and usual time/space coordinates, $(t,\vec{x})$, in Newtonian mechanics allowed us to define the quantum
propagator for massive scalar particles in a way very reminiscent to the computation of {\it{fixed-energy}} propagators in a non-relativistic
framework. The aforementioned analogy is viewed in many steps of the computation of the
semiclassical propagator, and particularly in the final result. We explicitly derived the first quantum corrections, encoded in the $F$ functional, to the classical path and showed that they keep
the form of a Van Wleck-Morette determinant with a $(g(\tau)g(0))^{1/4}$ prefactor, arising because of the space-time curvature. The
semiclassical dynamical equations are also computed and can be interpreted as the conservation of a four-current of probability, defined at the WKB
order as the product of the square modulus of quantum corrections times the classical four-momentum. Such a result is a direct generalization
of the semiclassical equations of motion obtained in any Schwarzshild-like background \cite{wkb}. Finally, as already mentioned in
\cite{hartle}, we prove that choosing an action functional quadratic in four-velocity is suitable for propagators as it leads to the
correct Feynman propagator in flat spaces.

	However, our computation opens new questions when compared to previous results. Inspired by non-relativistic quantum mechanics, the use
of the stationary phase approximation, to get results in the mass domain instead of the proper time domain, leads to a weight
different from the one considered in \cite{hartle}. The weight $w$ in \cite{hartle} was chosen by imposing the propagator to be a solution of the
inhomogeneous KG equation, assuming a multiplicative constant in the action functional equal to one quarter. In our study, the weight
is motivated by the stationary phase approximation, $w=\exp{(-\frac{i}{\hbar}\frac{\partial\tilde{S}}{\partial\tau}\tau)}$,
and the multiplicative constant entering the action functional is chosen so that the propagator is still a solution of the inhomogeneous KG equation.
These two different viewpoints to address the problem of {\it{general relativistic}} quantum mechanics as path integrals, are the fundamental
cause of the difference between our WKB propagator and the DeWitt's one \cite{dewitt}. We point out that imposing this stationary phase approximation, which is needed for
semiclassical studies, leads to a less ambiguous determination of the multiplicative constant $\alpha$ in the action functional than in \cite{hartle}. 
Furthermore, the Newtonian limit favors our proposal. We also suggest a new
experiment that could allow us to probe the quantum prefactor with interferences in the
gravitational field.
Anyway, this opens questions about the definition of the propagator which, to the best of our 
knowledge, have
never been considered, and should be addressed by studying quantum phenomena in curved space-times
which would allow to discriminate between the two definitions.

In addition to this ambiguity in the definition of the propagator, the potential-like curvature term involved in the equation of motion 
makes the comparison of the path integral formulation with the field theory approach more tricky. A complete comparison between quantum 
propagators derived from path integral with propagators derived from field theories will probably help building links between the two 
approaches.

\end{document}